# ARE A SET OF MICROARRAYS INDEPENDENT OF EACH OTHER?[1]

By Bradley Efron

*Stanford University*

Having observed an $m \times n$ matrix $X$ whose rows are possibly correlated, we wish to test the hypothesis that the columns are independent of each other. Our motivation comes from microarray studies, where the rows of $X$ record expression levels for $m$ different genes, often highly correlated, while the columns represent $n$ individual microarrays, presumably obtained independently. The presumption of independence underlies all the familiar permutation, cross-validation and bootstrap methods for microarray analysis, so it is important to know when independence fails. We develop nonparametric and normal-theory testing methods. The row and column correlations of $X$ interact with each other in a way that complicates test procedures, essentially by reducing the accuracy of the relevant estimators.

**1. Introduction.** The formal statistical problem considered here can be stated simply: having observed an $m \times n$ data matrix $X$ with possibly correlated rows, test the hypothesis that the columns are independent of each other. Relationships between the row correlations and column correlations of $X$ complicate the problem's solution.

Why are we interested in column-wise independence? The motivation in this paper comes from microarray studies, where $X$ is a matrix of expression levels for $m$ genes on $n$ microarrays. In the "Cardio" study I will use for illustration there are $m = 20,426$ genes each measured on $n = 63$ arrays, with the microarrays corresponding to 63 subjects, 44 healthy controls and 19 cardiovascular patients.[2] We expect the gene expressions to be correlated, inducing substantial correlations *within* each column [Owen (2005),

Received October 2008; revised October 2008.

[1]Supported in part by the NSF Grant DMS-00-72360 and by National Institute of Health Grant 8R01 EB002784.

[2]The entries of $X$ are log(red/green) ratios obtained from oligonucleotide arrays.

Efron (2007a), Qiu et al. (2005a)], but most of the standard analysis techniques begin with an assumption of independence *across* microarrays, that is, across the columns of $X$. This can be a risky assumption: all of the familiar permutation, cross-validation and bootstrap methods for microarray analysis, such as the popular SAM program of Tusher, Tibshirani and Chu (2001), depend on column-wise independence of $X$; dependence can invalidate the usual choice of a null hypothesis, as discussed next, leading to flawed assessments of significance.

An immediate purpose of the Cardio study is to identify genes involved in the disease process. For gene $i$ we compute the two-sample $t$-statistic "$t_i$" comparing sick versus healthy subjects. It will be convenient for discussion to convert these to $z$-scores,

$$(1.1) \qquad z_i = \Phi^{-1}(F_{61}(t_i)), \qquad i = 1, 2, \ldots, m,$$

with $\Phi$ and $F_{61}$ the cumulative distribution functions (c.d.f.) of standard normal and $t_{61}$ distributions; under the usual assumptions, $z_i$ will have a standard $N(0,1)$ null distribution, called here the "theoretical null." Unusually large values of $z_i$ or $-z_i$ are used to identify nonnull genes, with the meaning of "unusual" depending heavily on column-wise independence.

The left panel of Figure 1 shows the histogram of all 20,426 $z_i$ values, which is seen to be much wider than $N(0,1)$ near its center. An "empirical null" fit to the center, as in Efron (2007b), was estimated to be $N(0.03, 1.57^2)$. Null overdispersion has many possible causes [Efron (2004, 2007a, 2007b)], one of which is positive correlation across the columns of $X$. Such correlations reduce the effective degrees of freedom for the $t$-statistic, causing (1.1) to yield overdispersed null $z_i$s, and of course changing our assessment of significance for outlying values.

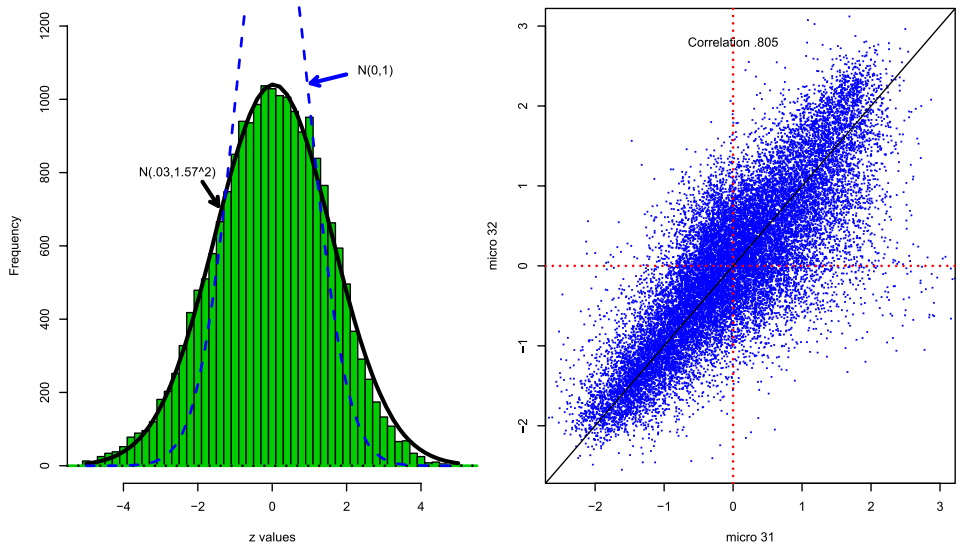

FIG. 1.   *Left panel: histogram of $m = 20{,}426$ z-values (1.1) for Cardio study; center of histogram is much wider than $N(0,1)$ theoretical null. Right panel: scatterplot of microarrays 31 and 32, $(x_{i31}, x_{i32})$ for $i = 1, 2, \ldots, m$, after removal of row-wise gene means; the scattergram seems to indicate substantial correlation between the two arrays.*

The right panel of Figure 1 seems to offer a "smoking gun" for correlation: the scattergram of expression levels for microarrays 31 and 32 looks strikingly correlated, with sample correlation coefficient 0.805. Here $X$ has been standardized by subtraction of its row means, so the effect is not due to so-called ecological correlations. ($X$ is actually "doubly standardized," as defined in Section 2). Nevertheless, the question of whether or not correlation 0.805 is significantly positive turns out to be surprisingly close, as discussed in Section 4, because the row-wise correlations in $X$ drastically reduce the degrees of freedom for the scatterplot. Despite the massive appearance of 20,426 points, the scattergram's accuracy is no more than would be given by 17 independent bivariate normal pairs.

Answering the title's question, that is, testing for column-wise independence in the presence of row-wise dependence, has both easy and difficult aspects. Section 2 introduces a class of simple permutation tests which, in the case of the Cardio data, clearly discredit column-wise independence. However, these tests depend on the ordering of the $n$ columns, and can't be used if the initial order is lost. It is natural and desirable to look for test statistics of column-wise independence that are invariant under permutation of the columns. Classical multivariate analysis, as in Anderson (2003), develops column independence tests in terms of the eigenvalues of an $n$ by $n$ Wishart matrix. However, this theory depends on the assumption of row-wise independence, disqualifying it for use here.

Sections 3 through 5 consider more general classes of independence tests, both from nonparametric and normal theory points of view. The theorem in Section 3 illustrates a key difficulty: correlation between the rows of $X$ (ruled out in the classic theory) can give a misleading appearance of column-wise dependence. Similarly, row-wise dependence can greatly degrade the accuracy of the usual $n \times n$ sample covariance matrix of the columns, as shown by the theorem in Section 4. Various nonpermutation normal-theory tests are discussed in Section 5, some promising, but with difficulties seen for all of them. The paper ends in Section 6 with a collection of remarks and details.

**2. Permutation tests of column-wise independence.** Simple permutation tests can provide strong evidence against column-wise independence, as we will see for the Cardio data. Our main example concerns the 44 healthy subjects, where $X$ is now an $m \times n$ matrix with $m = 20{,}426$ and $n = 44$. For convenience, we assume that $X$ has been "demeaned" by the subtraction of row and column means, giving

$$(2.1) \qquad \sum_i x_{ij} = \sum_j x_{ij} = 0 \qquad \text{for } i = 1, 2, \ldots, m \text{ and } j = 1, 2, \ldots, n.$$

Our numerical results go further and assume "double standardization": that in addition to (2.1),

$$(2.2) \qquad \sum_j x_{ij}^2 = n \quad \text{and} \quad \sum_i x_{ij}^2 = m \qquad \text{for } i = 1, \ldots, m \text{ and } j = 1, \ldots, n,$$

that is, that each row and column of $X$ has mean 0 and variance 1; see Remark 6.4 in Section 6.

Let $\widehat{\Delta}$ be the familiar estimate of the $n \times n$ covariance matrix $\Delta$ between the columns of $X$,

$$(2.3) \qquad \widehat{\Delta} = (X'X)/m.$$

Under double standardization, $\widehat{\Delta}$ is actually the sample correlation matrix, which we expect to be near the identity matrix $I_n$ under column-wise independence. Also let $v_1$ denote the first eigenvector of $\widehat{\Delta}$. The left panel of Figure 2 plots the components of $v_1$ versus array number $1, 2, \ldots, 44$. Suppose that the columns of the original expression matrix, before standardization, are independent and identically distributed $m$-vectors ("i.i.d."). Then it is easy to see (Remark 6.2 of Section 6) that all orderings of the components of $v_1$ are equally likely. This is not what Figure 2 shows: the components seem to increase from left to right, with a noticeable block of large values for arrays 27–32.

Let $S(v_1)$ be a statistic that measures structure, for instance, a linear regression of $v_1$ versus array index. Comparing $S(v_1)$ with a set of permuted values

$$(2.4) \qquad \{S^{*l} = S(v^{*l}), \ l = 1, 2, \ldots, L\},$$

$v^{*l}$ a random permutation of the components of $v_1$, provides a quick test of the i.i.d. null hypothesis.

Permutation testing was applied to $v_1$ for the Cardio data, using the "block" statistic

$$(2.5) \qquad S(v_1) = v_1' B v_1,$$

where $B$ is the $n \times n$ matrix

$$(2.6) \qquad B = \sum_h \beta_h \beta_h'.$$

The sum in (2.6) is over all vectors $\beta_h$ of the form

$$(2.7) \qquad \beta_h = (0, 0, \ldots, 0, 1, 1, \ldots, 1, 0, 0, \ldots, 0),$$

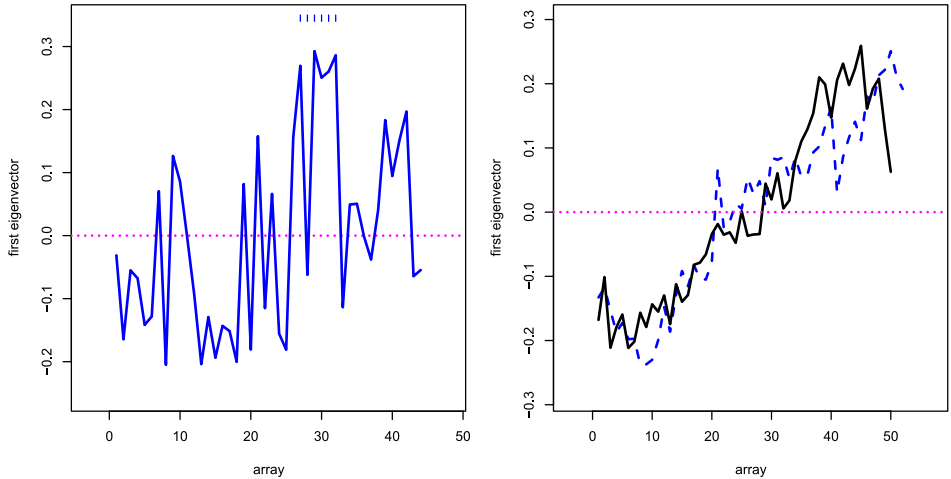

FIG. 2.  *Left panel: Components of first eigenvector of row sample correlation matrix for the 44 healthy Cardio subjects, plotted versus array number* $1, 2, \ldots, 44$; *dashes emphasize the block of large components for arrays* 27–32. *Right panel: First eigenvectors for healthy (solid line) and cancer (dashed) subjects, prostate cancer study, Singh et al. (2002); there was a systematic drift in expression levels as the study progressed.*

with the 1s forming blocks of length between 2 and 10 inclusive. A heuristic rationale for block testing appears below; intuitively, microarray experiments are prone to block disturbances because of the way they are developed and read; see Callow et al. (2000). After $L = 5000$ permutations, only three $S^*$ values exceeded the actual value $S(v_1)$, $p$-value 0.0006, yielding strong evidence against the i.i.d. null hypothesis.

The right panel of Figure 2 pertains to a microarray prostate cancer study [Singh et al. (2002)] discussed in Efron (2008): $m = 6033$ genes were measured on each of $n = 102$ men, 50 healthy controls and 52 prostate cancer patients. The right panel plots first eigenvectors for $\widehat{\Delta}$, (2.3), computed separately for the healthy controls and the cancer patients (the two matrices being individually doubly standardized). Both vectors increase almost linearly from left to right. Taking $S(v_1)$ as the linear regression of $v_1$ versus array number, permutation testing overwhelmingly rejected the i.i.d. null hypothesis, as it also did using the block test. The prostate study appears as a favorable example of microarray technology in Efron (2008). Nevertheless, Figure 2 indicates a systematic drift in the expression level readings as the study progressed. Some genes drift up, others down (the average drift equaling 0 because of standardization), inducing a small amount of column-wise correlation.

Section 5 discusses models for $X$ where the $n \times n$ column covariance matrix $\Delta$ is of the "single degree of freedom" form

$$(2.8) \qquad \Delta = I + \lambda \beta \beta'$$

for some known fixed vector $\beta$, the null hypothesis of column-wise independence being $H_0 : \lambda = 0$. An obvious choice of test statistic in this situation is

$$(2.9) \qquad S_\beta = \beta'(\widehat{\Delta} - I)\beta,$$

a monotone increasing function of $\beta'\widehat{\Delta}\beta$. If $\beta$ is unknown, we can replace $S_\beta$ with

$$(2.10) \qquad S_B = \sum_{h=1}^{H} \beta_h' \widehat{\Delta} \beta_h = \operatorname{tr}\left(\widehat{\Delta} \sum_h \beta_h \beta_h'\right) \equiv \operatorname{tr}(\widehat{\Delta} B),$$

where $\{\beta_1, \beta_2, \ldots, \beta_H\}$ is a catalog of "likely prospects" as in (2.7).

Permutation test statistics such as (2.5) can be motivated from the singular value decomposition (SVD) of $X$,

$$(2.11) \qquad \underset{m \times n}{X} = \underset{m \times K}{U} \underset{K \times K}{d} \underset{K \times n}{V'},$$

where $K$ is the rank, $d$ the diagonal matrix of ordered singular values, and $U$ and $V$ orthonormal matrices of sizes $m \times K$ and $n \times K$,

$$(2.12) \qquad U'U = V'V = I_K,$$

$I_K$ the $K \times K$ identity. The squares of the diagonal elements, say,

$$(2.13) \qquad e_1 \geq e_2 \geq \cdots \geq e_K > 0 \qquad (e_k = d_k^2),$$

are the eigenvalues of $X'X = V'd^2 V$.

$S_B$ in (2.10) can now be written as

$$(2.14) \qquad S_B = \sum_{j=1}^{k} \frac{e_j}{m}(v_j' B v_j).$$

Model (2.8) suggests that most of the information against the null hypothesis $H_0$ of independence lies in the first eigenvector $v_1$, getting us back to test statistic $S(v_1) = v_1' B v_1$ as in (2.5).

What should the statistician do if column-wise independence is strongly rejected, as in the Cardio example? Use of an empirical null rather than a permutation or theoretical null, $\mathcal{N}(0.03, 1.57^2)$ rather than $\mathcal{N}(0, 1)$ in Figure 1, removes the reliance on column-wise independence for hypothesis testing methods such as False Discovery Rates, at the expense of increased variability. Efron (2008) discusses these points.

Two objections can be raised to our permutation tests: (1) they are really testing i.i.d., not independence; (2) nonindependence might not manifest itself in the order of $v_1$ (particularly if the order of the microarrays has been shuffled in some unknown way).

Column-wise standardization makes the column distributions more similar, mitigating objection (1). Going further, "quantile standardization"— say, replacing each column's entries by normal scores [Bolstad et al. (2003)]— makes the marginals exactly the same. The Cardio data was reanalyzed using normal scores, with almost identical results.

Objection (2) is more worrisome from the point of view of statistical power. The order in which the arrays were obtained *should* be available to the statistician, and should be analyzed to expose possible trends like those in Figure 2.[3] It would be desirable, nevertheless, to have independence tests that do not depend on order—that is, test statistics invariant under column-wise permutations. The remainder of this paper concerns both the possibilities and difficulties in the development of "nonpermutation" tests.

**3. Row and column correlations.** There is an interesting relationship between the row and column correlations of the matrix $X$, which complicates the question of column-wise independence. For the notation of this section define the $n \times n$ matrix of sample covariances between the columns of $X$ as

$$(3.1) \qquad \widehat{\mathbf{Cov}} = X'X/m,$$

called $\widehat{\Delta}$ in Section 2, and likewise

$$(3.2) \qquad \widehat{\mathbf{cov}} = XX'/n,$$

for the $m \times m$ matrix of row-wise sample covariances (having more than 400,000,000 entries in the Cardio example!).

THEOREM 1.  *If $X$ has row and column means* 0, *(2.1), then the $n^2$ entries of $\widehat{\mathbf{Cov}}$ have empirical mean* 0 *and variance $c_2$,*

$$(3.3) \qquad c_2 = \sum_{k=1}^{K} e_k^2/(mn)^2,$$

*with $e_k$ the eigenvalues (2.13), and so do the $m^2$ entries of $\widehat{\mathbf{cov}}$.*

PROOF.  The sum of $\widehat{\mathbf{Cov}}$'s entries is

$$(3.4) \qquad 1_n' X'X 1_n/m = 0,$$

---

[3]The referee points out that when Affymetrix CEL files are available, array run dates will usually be found in the `DatHeader` lines.

according to (2.1), while the mean of squared entries is

$$(3.5) \qquad \frac{\sum_{j=1}^{n} \sum_{j'=1}^{n} \widehat{Cov}_{jj'}^2}{n^2} = \frac{\mathrm{tr}((X'X)^2)}{m^2 n^2} = \frac{\mathrm{tr}(V'd^4 V)}{m^2 n^2} = c_2.$$

Replacing $X'X$ with $XX'$ yields the same results for the row covariances $\widehat{\mathbf{cov}}$. $\quad\square$

Under double standardization (2.1)–(2.2), the covariances become sample correlations, say, $\widehat{\mathbf{Cor}}$ and $\widehat{\mathbf{cor}}$ for the columns and rows. Theorem 1 has a surprising consequence: *whether or not the columns of $X$ are independent, the column sample correlations will have the same mean and variance as the row correlations.* In other words, substantial row-wise correlation can induce the appearance of column-wise correlation.

Figure 3 concerns the 44 healthy subjects in the Cardio study, with $X$ an $(m, n) = (20{,}426{,}44)$ doubly standardized matrix. All $44^2$ column correlations are shown by the solid histogram, while the line histogram is a random sample of 10,000 row correlations. Here $c_2 = 0.283^2$, so according to the theorem, both histograms have mean 0 and standard deviation 0.283.

The 44 diagonal elements of $\widehat{\mathbf{Cor}}$ protrude as a prominent spike at 1. (We can not see the spike of 20,426 diagonal elements for the row correlation matrix $\widehat{\mathbf{cor}}$ because they form such a small fraction of all $20{,}426^2$.) It is easy to remove the diagonal 1's from consideration.

COROLLARY. *In the doubly standardized situation, the off-diagonal elements of the column correlation matrix $\widehat{\mathbf{Cor}}$ have empirical mean and variance*

$$(3.6) \qquad \hat{\mu} = -\frac{1}{n-1} \quad and \quad \hat{\alpha}^2 = \frac{n}{n-1}\left(c_2 - \frac{1}{n-1}\right).$$

For $n = 44$ and $c_2 = 0.283$ this gives

$$(3.7) \qquad (\hat{\mu}, \hat{\alpha}^2) = (-0.023, 0.241^2).$$

The corresponding diagonal-removing corrections for the row correlations [replacing $n$ by $m$ in (3.6)] are negligible for $m = 20{,}426$. However, $c_2$ overestimates the variance of the row correlations for another reason: with only 44 points available to estimate each correlation, estimation error adds a considerable component of variance to the $\widehat{\mathbf{cor}}$ histogram in the left panel, as discussed next.

Suppose now that the columns of $X$ are in fact independent, in which case the substantial column correlations seen in Figure 3 must actually be induced by row correlations, via Theorem 1. Let $cor_{ii'}$ indicate the *true* correlation between rows $i$ and $i'$ (i.e., between $X_{ij}$ and $X_{i'j}$), and define $\alpha$ the *total correlation* to be the root mean square of the $cor_{ii'}$ values,

$$(3.8) \qquad \alpha^2 = \sum_{i<i'} cor^2_{ii'} \Big/ \binom{m}{2}.$$

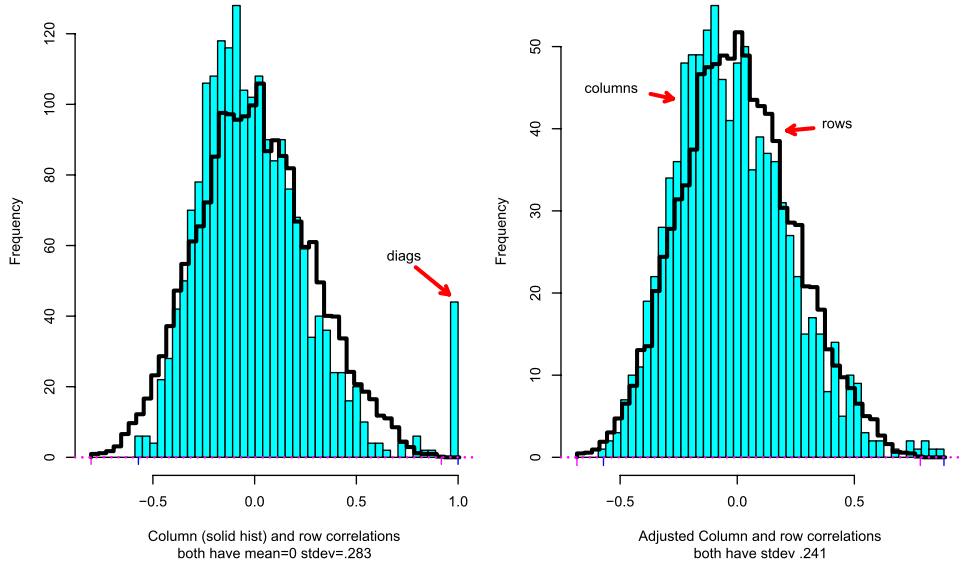

Fig. 3.   *Left panel: solid histogram the $44^2$ column sample correlations for $X$ the doubly standardized matrix of healthy Cardio subjects; line histogram is sample of 10,000 of the $20,426^2$ row correlations. Right panel: solid histogram the column correlations excluding diagonal 1s; line histogram the row correlations corrected for sampling overdispersion.*

Remark 6.5 of Section 6 shows that $\hat\alpha^2$ in (3.6) is an approximately unbiased estimate of $\alpha^2$, assuming column-wise independence. For the Cardio example $\hat\alpha = 0.241$, similar to the size of the microarray correlation estimates in Efron (2007a), Owen (2005) and Qiu et al. (2005a). Section 4 discusses the crucial role of $\alpha$ in determining the accuracy of estimates based on $X$.

The right panel of Figure 3 compares the histogram of the column correlations $\widehat{Cor}_{jj'}$, now excluding cases $j = j'$, with the row correlation histogram corrected for sampling overdispersion via the shrinkage factor $0.0241/0.283$. As predicted by Theorem 1, the similarity is striking. A possible difference lies in the long right tail of the $\widehat{\mathbf{Cor}}$ distribution (including $\widehat{Cor}_{31,32}$, the case illustrated in Figure 1), whose significance is examined in Section 4.

**4. Normal theory.** The results of Sections 2 and 3 were developed non-parametrically. This section concerns multivariate normal theory, afterward used in Section 5 to draw the connection with classical multivariate independence tests. We consider the *matrix normal* distribution for $X$,

$$(4.1) \qquad \underset{m \times n}{X} \sim \mathcal{N}_{m,n}(0, \underset{m \times m}{\Sigma} \otimes \underset{n \times n}{\Delta}),$$

where the Kronecker notation indicates covariance structure

$$(4.2) \qquad \text{cov}(X_{ij}, X_{i'j'}) = \Sigma_{ii'} \Delta_{jj'}.$$

Row $x_i$ of $X$ has covariance matrix proportional to $\Delta$,

$$(4.3) \qquad x_i \sim \mathcal{N}_n(0, \Sigma_{ii} \Delta)$$

(*not* independently across rows unless $\Sigma$ is diagonal), and likewise for column $\mathbf{x}_j$, $\mathbf{x}_j \sim \mathcal{N}_m(0, \Delta_{jj} \Sigma)$. As in (2.1), we take all means equal to 0.

Much of classical multivariate analysis focuses on the situation $\Sigma = I$, where the rows $x_i$ are independent replicates,[4]

$$(4.4) \qquad \Sigma = I : x_i \overset{\text{i.i.d.}}{\sim} \mathcal{N}_n(0, \Delta), \qquad i = 1, 2, \ldots, m,$$

in which case the sample covariance matrix $\widehat{\Delta} = X'X/m$ has a scaled Wishart distribution,

$$(4.5) \qquad \widehat{\Delta} \sim \text{Wishart}(m, \Delta)/m.$$

Distribution (4.5) has first and second moments

$$(4.6) \qquad \underset{n \times n}{\widehat{\Delta}} \sim (\underset{n \times n}{\Delta}, \underset{n^2 \times n^2}{\Delta^{(2)}}/m) \qquad \text{with } \Delta_{jk,lh}^{(2)} = \Delta_{jl}\Delta_{kh} + \Delta_{jh}\Delta_{kl}$$

for $j, k, l, h = 1, 2, \ldots, n$; see Mardia, Kent and Bibby [(1979), page 92].

Relation (4.6) says that when $\Sigma = I$, that is when the rows of $X$ are independent, $\widehat{\Delta}$ unbiasedly estimates the row covariance matrix $\Delta$ with accuracy proportional to $m^{-1/2}$. Correlation between rows reduces the accuracy of $\widehat{\Delta}$, as shown next.

Returning to the general situation (4.1)–(4.3), define

$$(4.7) \qquad \widetilde{\Delta} = X'\boldsymbol{\sigma}^{-2}X/m,$$

where $\boldsymbol{\sigma}$ is the diagonal matrix with diagonal entries $\sigma_i = \Sigma_{ii}^{1/2}$.

THEOREM 2. *Under model (4.1), $\widetilde{\Delta}$ has first and second moments*

$$(4.8) \qquad \widetilde{\Delta} \sim (\Delta, \Delta^{(2)}/\tilde{m}), \qquad \tilde{m} = m/[1 + (m-1)\alpha^2],$$

---

[4] Most multivariate texts reverse the situation, taking the columns as independent replicas of possibly correlated rows.

*where $\alpha$ is the total correlation as in (3.8),*

$$(4.9) \qquad \alpha^2 = \sum_{i<i'} (\Sigma_{ii'}^2 / \Sigma_{ii} \Sigma_{i'i'}) \Big/ \binom{m}{2},$$

*and $\Delta^{(2)}$ is the Wishart covariance (4.6).*

Comparing (4.8) with (4.6), we see that correlation between the rows reduces "effective sample size" from $m$ to $\tilde{m}$: for $\alpha = 0.241$ as in (3.7), the reduction is from $m = 20{,}426$ to $\tilde{m} = 17.2$! (Notice that row standardization effectively makes $\sigma_i \doteq 1$ in (4.7), so $\tilde{\Delta} \doteq \widehat{\Delta}$ (2.3), justifying the comparison.) The total correlation $\alpha$ shows up in other efficiency calculations; see Remark 6.7.

PROOF OF THEOREM 2. The row-standardized matrix $\tilde{X} = \boldsymbol{\sigma}^{-1} X$ has matrix normal distribution

$$(4.10) \qquad \tilde{X} \sim \mathcal{N}_{m,n}(0, \tilde{\Sigma} \otimes \Delta),$$

where $\tilde{\Sigma} = \boldsymbol{\sigma}^{-1} \Sigma \boldsymbol{\sigma}^{-1}$ has diagonal elements $\tilde{\Sigma}_{ii} = 1$. From (4.2) we see that $\tilde{\Sigma}_{ii'} = \Sigma_{ii'} / (\Sigma_{ii} \Sigma_{i'i'})^{1/2}$ is the correlation between elements $X_{ij}$ and $X_{i'j}$ in the same column of $X$; $\tilde{\Delta} = \tilde{X}' \tilde{X} / m$ has entries $\tilde{\Delta}_{jk} = \sum_i \tilde{X}_{ij} \tilde{X}_{ik} / m$, and is unbiased for $\Delta$,

$$(4.11) \qquad E\{\tilde{\Delta}_{jk}\} = \Delta_{jk},$$

using (4.2).

The covariance calculation for $\tilde{\Delta}$ involves expansion

$$(4.12) \qquad \tilde{\Delta}_{jk} \tilde{\Delta}_{lh} = \left( \sum_i \tilde{X}_{ij} \tilde{X}_{ik} / m \right) \left( \sum_{i'} \tilde{X}_{i'l} \tilde{X}_{i'h} / m \right)$$

$$(4.13) \qquad = \frac{1}{m^2} \left( \sum_i \tilde{X}_{ij} \tilde{X}_{ik} \tilde{X}_{il} \tilde{X}_{ih} + \sum_{i \neq i'} \tilde{X}_{ij} \tilde{X}_{ik} \tilde{X}_{i'l} \tilde{X}_{i'h} \right).$$

Using the formula

$$(4.14) \qquad E\{Z_1 Z_2 Z_3 Z_4\} = \gamma_{12} \gamma_{34} + \gamma_{13} \gamma_{24} + \gamma_{14} \gamma_{23}$$

for a normal vector $(Z_1 Z_2 Z_3 Z_4)'$ with 0 means and covariances $\gamma_{ij}$, (4.2) gives

$$(4.15) \qquad E\left\{ \sum_i \tilde{X}_{ij} \tilde{X}_{ik} \tilde{X}_{il} \tilde{X}_{ih} \right\} = m[\Delta_{jk} \Delta_{lh} + \Delta_{jl} \Delta_{kh} + \Delta_{jh} \Delta_{kl}]$$

and

$$E\left\{\sum_{i\neq i'}\widetilde{X}_{ij}\widetilde{X}_{ik}\widetilde{X}_{i'l}\widetilde{X}_{i'h}\right\} = m(m-1)\Delta_{jk}\Delta_{lh}$$

(4.16)

$$+ (\Delta_{jl}\Delta_{kh} + \Delta_{jh}\Delta_{kl})\sum_{i\neq i'}\widetilde{\Sigma}_{ii'}^2.$$

Then (4.13) yields

$$(4.17) \quad E\{\widetilde{\Delta}_{jk}\widetilde{\Delta}_{lh}\} = \Delta_{jk}\Delta_{lh} + (\Delta_{jl}\Delta_{kh} + \Delta_{jh}\Delta_{kl})\left(\frac{1+(m-1)\alpha^2}{m}\right),$$

giving

$$(4.18) \qquad \mathrm{cov}(\widetilde{\Delta}_{jk}, \widetilde{\Delta}_{lh}) = (\Delta_{jl}\Delta_{kh} + \Delta_{jh}\Delta_{kl})/\tilde{m},$$

as in (4.8).    □

A corollary of Theorem 2, used in Section 5, concerns bilinear functions of $\Delta$ and $\widetilde{\Delta}$,

$$(4.19) \qquad \tau^2 = w'\Delta w \quad \text{and} \quad \tilde{\tau}^2 = w'\widetilde{\Delta}w,$$

where $w$ is a given $n$-vector.

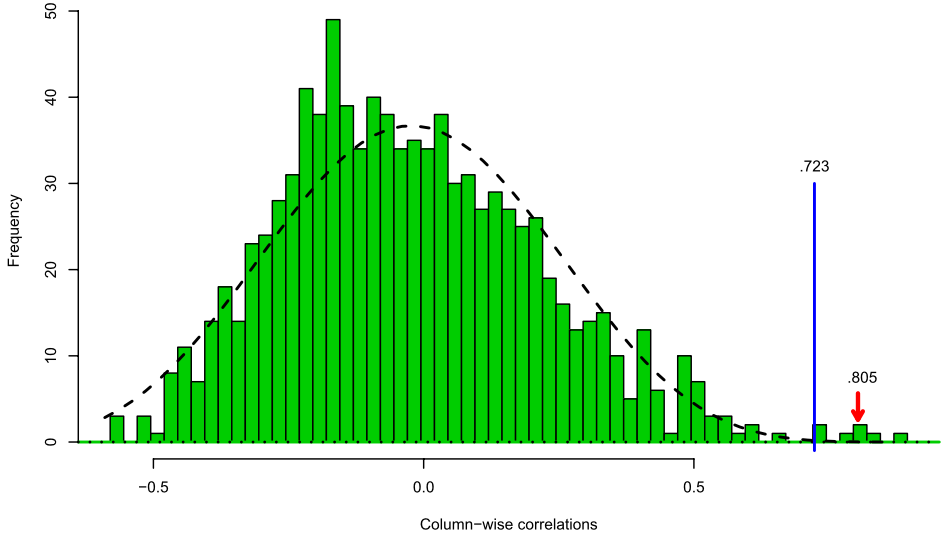

Fig. 4. *Dashed curve is normal-theory null density for correlation coefficient from $\bar{m} = 17.2$ pairs of points; see Remark 6.6. Histogram is the 946 column correlations, right panel Figure 3. FDR test, $q = 0.1$, yielded 7 significant correlations, $\widehat{Cor} \geq 0.723$, including 0.805 between arrays 31 and 32, Figure 1.*

COROLLARY.   *Under model (4.1), $\tilde{\tau}^2$ has mean and variance*

$$(4.20) \qquad \tilde{\tau}^2 \sim (\tau^2, 2\tau^4/\tilde{m}).$$

The proof follows that for Theorem 2; see Remark 6.9.

If $\Sigma = I$ in (4.1), then $\tilde{\Delta} = \hat{\Delta}$ and $\tilde{\tau}^2$ has a scaled chi-squared distribution,

$$(4.21) \qquad \tilde{\tau}^2 \sim \tau^2 \cdot \chi_m^2/m,$$

with mean and variance $\tilde{\tau}^2 \sim (\tau^2, 2\tau^4/m)$, so again the effect of correlation within $\Sigma$ is to reduce the effective sample size from $m$ to $\tilde{m}$ (4.8).

We can approximate $\tilde{\Delta}$ (4.7), with

$$(4.22) \qquad \hat{\Delta} = X' \hat{\boldsymbol{\sigma}}^{-2} X/m,$$

where $\hat{\sigma}_{ii}^2$ is an estimate of $\Sigma_{ii}$ based on the observed variability in row $i$. If the rows of $X$ have been standardized, then $\hat{\sigma}_{ii}^2 = 1$ and $\hat{\Delta}$ returns to its original definition $X'X/m$.

Both Theorem 2 and the Corollary encourage us to think of $\hat{\Delta}$ as, approximately, a scaled Wishart distribution based on an independent sample of size $\tilde{m}$,

$$(4.23) \qquad \hat{\Delta} \sim \text{Wishart}(\tilde{m}, \Delta)/\tilde{m}.$$

The dangers of this approximation are discussed in Section 5, but it is, nevertheless, an evocative heuristic, as shown below.

Figure 4 returns to the question of the seemingly overwhelming correlation 0.805 between arrays 31 and 32 seen in Figure 1. A one-sided $p$-value was calculated for each of the 946 column correlations, using as a null hypothesis the normal theory correlation coefficient distribution based on a sample size of $\tilde{m} = 17.2$ pairs of $\mathcal{N}_2(0, I)$ points [the correct null if $\Delta = I$ in (4.23)]. Benjamini and Hochberg's (1995) False Discovery Rate test, level $q = 0.1$, was applied to the 946 $p$-values. This yielded 7 significant cases, those with sample correlation $\geq 0.723$; all 7 were from the block of arrays 27 to 32 indicated in Figure 2. Correlation 0.805 does turn out to be significant, but by a much closer margin than Figure 1's scattergram suggests.

The FDR procedure was also applied using the simpler null distribution $\mathcal{N}(-0.023, 0.241^2)$ (3.7). This raised the significance threshold from 0.723 to 0.780, removing two of the previously significant correlations.

Theorem 1 showed that the variance of the observed column correlations is useless for testing column-wise independence, since any value at all can be induced by row correlations. The test in Figure 4 avoids this trap by looking for unusual outliers among the column correlations. It does *not* depend on the order of the columns, objection (2) in Section 2 for permutation tests, but pays the price of increased modeling assumptions.

**5. Other test statistics.** Theorem 2 offers a normal-theory strategy for testing column-wise independence. We begin with $X \sim \mathcal{N}_{m,n}(0, \Sigma \otimes \Delta)$ (4.1), taking

$$(5.1) \qquad \Sigma_{ii} = 1 \quad \text{and} \quad \Delta_{jj} = 1 \qquad \text{for all } i \text{ and } j,$$

as suggested by double standardization. The null hypothesis of column-wise independence is equivalent to the column correlation matrix $\Delta$ equaling the identity,

$$(5.2) \qquad H_0 : \Delta = I,$$

since then (4.2) says that all pairs in different columns are independent.

To test (5.2), we estimate $\Delta$ with $\widehat{\Delta}$, (4.22) or more simply $\widehat{\Delta} = X'X/m$ after standardization, and compute a test statistic

$$(5.3) \qquad S = s(\widehat{\Delta}),$$

where $s(\cdot)$ is some measure of distance between $\widehat{\Delta}$ and $I$. The accuracy approximation $\widehat{\Delta} \overset{.}{\sim} (\Delta, \Delta^{(2)}/\tilde{m})$ from (4.8), with $\Delta = I$, is used to assess the significance level of the observed $S$, maybe even employing the more daring approximation $\widehat{\Delta} \overset{.}{\sim} \text{Wishart}(\tilde{m}, I)/\tilde{m}$. Strategy (5.3) looks promising but, as the examples of this section will show, it suffers from serious difficulties that are absent under the classic assumption of independent rows.

One of the difficulties stems from Theorem 1. An obvious test statistic for $H_0 : \Delta = I$ is

$$(5.4) \qquad S = \sum_{j < j'} \widehat{\Delta}_{j,j'}^2 \Big/ \binom{n}{2},$$

the average squared off-diagonal element of $\widehat{\Delta}$. But $\widehat{\Delta} = \widehat{\mathbf{Cov}}$ (3.1), so in the doubly standardized situation of (3.6), $S$ is an increasing monotone function of $\hat{\alpha}$, the estimated total correlation. This disqualifies $S$ as a test statistic for (5.2), since large values of $\hat{\alpha}$ can always be attributed to row-wise correlation alone.

Similarly, the variance of the eigenvalues (2.13),

$$(5.5) \qquad S = \sum_{k=1}^{K} (e_k - e_{\cdot})^2 / k \qquad \Big( e_{\cdot} = \sum e_k / K \Big),$$

looks appealing since the true eigenvalues all equal 1 when $\Delta = I$. However, (5.5) is also a monotonic function of $\hat{\alpha}$; see Remark 6.1.

The general difficulty here is "leakage," the fact that row-wise correlations affect the observed pattern of column-wise correlations. This becomes clearer by comparison with classical multivariate methods, where row-wise correlations are assumed away by taking $\Sigma = I$ in (4.1). Johnson and Graybill

([1972](#)) consider a two-way ANOVA problem where, after subtraction of main effects, $X$ has the form

$$(5.6) \qquad X_{ij} = a_i\beta_j + \varepsilon_{ij} \qquad \text{for } i = 1, 2, \ldots, m \text{ and } j = 1, 2, \ldots, n,$$

$a_i \sim \mathcal{N}(0, \lambda)$ and $\varepsilon_{ij} \sim \mathcal{N}(0, 1)$, all independently, with $\beta = (\beta_1, \beta_2, \ldots, \beta_n)$ a fixed but unknown vector (representing "one degree of freedom for non-additivity" in the two-way table $X$, Johnson and Graybill's extension of Tukey's procedure).

In the Kronecker notation (4.1), $X \sim \mathcal{N}_{m,n}(0, I \otimes \Delta)$ with

$$(5.7) \qquad \Delta = I + \lambda\beta\beta'.$$

Now (5.2) becomes $H_0 : \lambda = 0$. Johnson and Graybill show that, with $\beta$ unknown, the likelihood ratio test rejects $H_0$ for large values of the eigenvalue ratio (2.13),

$$(5.8) \qquad S = e_1 \Big/ \sum_{k=1}^{K} e_k.$$

Since the $m$ rows of $X$ are assumed independent, they can test $H_0$ by comparison of $S$ with values $S^* = e_1^* / \sum_{k=1}^{K} e_k^*$ obtained from

$$(5.9) \qquad \widehat{\Delta}^* \sim \text{Wishart}(m, I)/m,$$

as in (4.5).

Getting back to the correlated rows situation, Theorem 2 suggests comparing $S$ with values $S^*$ from

$$(5.10) \qquad \widehat{\Delta}^* \sim \text{Wishart}(\tilde{m}, I)/\tilde{m},$$

$\tilde{m}$ as in (4.8). The solid histogram in Figure 5 compares 100 $S^*$ values from (5.10), $\tilde{m} = 17.2$ for the Cardio data, with the observed value $S = 0.207$ from the doubly standardized Cardio matrix for the healthy subjects used in Figure 3. All 100 $S^*$ values are much smaller than $S$, providing strong evidence against $H_0 : \Delta = I$.

The evidence looks somewhat weaker, though, if we simulate $S^*$ values with $\widehat{\Delta}^*$ obtained from random matrices

$$(5.11) \qquad X^* \sim \mathcal{N}_{20,426,44}(0, \widehat{\Sigma} \otimes I),$$

doubly standardized, where $\widehat{\Sigma}$ has total correlation $\alpha = 0.241$, the estimated value for $X$, (4.9). The line histogram in Figure 5 shows 100 such $S^*$ values, all still smaller than $S$, but substantially less so. (Remark 6.8 describes the construction of $X^*$.)

Why does (5.11) produce larger "null" $S^*$ values than (5.10)? The answer is simple: even though the first and second moments of $\widehat{\Delta}^* = \boldsymbol{X}^{*\prime}\boldsymbol{X}^*/m$

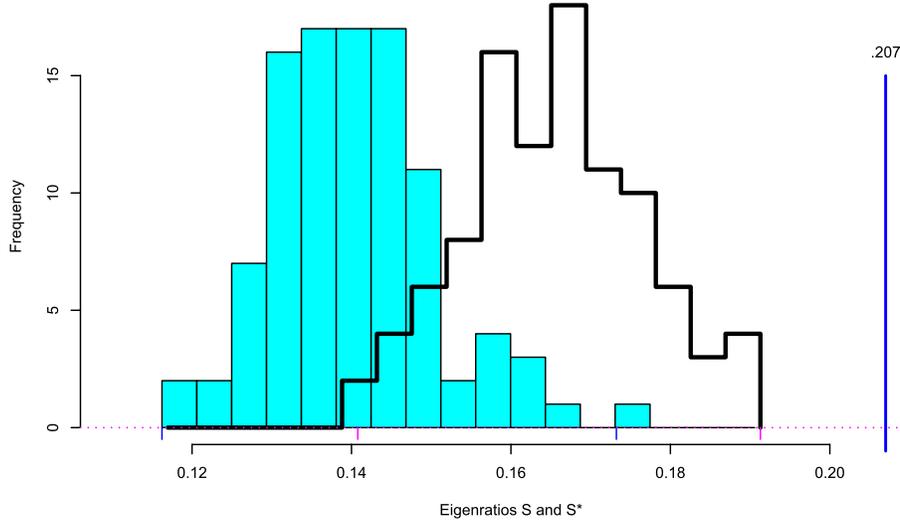

FIG. 5.  *Eigenratio statistic (5.8) equals* 0.207 *for 20,426 × 44 Cardio matrix X; solid histogram* 100 *simulations* $S^*$ *from Wishart (5.10),* $\tilde{m} = 17.2$*; line histogram* 100 *simulations from correlated-row* $X^*$ *matrices (5.11),* $\alpha = 0.241, \Delta = I$*.*

match $\widehat{\Delta}^*$ from (5.10), its eigenvalues do not. The nonzero eigenvalues of $X^{*\prime}X^*/m$ equal those of $\widehat{\Sigma}^* = X^*X^{*\prime}/n$. This is another example of leakage, where the fact that $\Sigma$ in (5.11) is not the identity $I_m$ distorts the estimated eigenvalue of $\widehat{\Delta}^*$, even if $\Delta = I_n$.

The eigenratio statistic $S = e_1 / \sum e_k$ is invariant under permutations of the columns of $X$, answering objection (2) to permutation testing of Section 2. Because of invariance, the eigenratio and permutation tests provide *independent* $p$-values for testing the null hypothesis of i.i.d. columns, and so can be employed together. Figure 5 is disturbing nonetheless, in suggesting that an appropriate null distribution for $S$ depends considerably on the choice of the nuisance parameter $\Sigma$ in (5.11).

The bilinear form (4.19)–(4.20) yields another class of test statistics,

$$\hat{\tau}^2 = w'\widehat{\Delta}w \,\dot{\sim}\, (\tau^2, 2\tau^4/\tilde{m}), \tag{5.12}$$

where $w$ is a pre-chosen $n$-vector and $\tau^2 = w'\Delta w$. Delta-method arguments give $\mathrm{CV}(\hat{\tau}) \doteq (2\tilde{m})^{-1/2}$ for the coefficient of variation of $\hat{\tau}$. Defining

$$Z_i = x_i'w \qquad (x_i' \text{ the } i\text{th row of } X) \tag{5.13}$$

yields the alternative form

$$\hat{\tau}^2 = \sum_{i=1}^{m} Z_i^2/m. \tag{5.14}$$

In a two-sample situation like that for the Cardio study, sample sizes $n_1$ and $n_2$, we can choose

$$(5.15) \qquad w' = \left(\frac{n_1 n_2}{n_1 + n_2}\right)^{1/2} (-1_{n_1}/n_1, 1_{n_2}/n_2),$$

"$1_n$" indicating a vector of $n$ 1's. This choice makes

$$(5.16) \qquad Z_i = \left(\frac{n_1 n_2}{n_1 + n_2}\right)^{1/2} (\bar{x}_{2i} - \bar{x}_{1i}),$$

the multiple of the mean response difference between the two samples that has variance 1 if $\Delta = I$. In terms of (5.12), $\|w\|^2 = 1$ so $\tau^2 = 1$.

For the Cardio study, with $n_1 = 44, n_2 = 19$, and $\tilde{m} = 17.2$, we obtain $\hat{\tau} = 1.48$, coefficient of variation 0.17. This puts $\hat{\tau}$ more than 2.8 standard errors above the null hypothesis value $\tau = 1$, again providing evidence against column-wise independence. The $Z_i$ values from (5.16) are nearly indistinguishable from the $z_i$ values in Figure 1—not surprisingly since with the rows of $X$ standardized, $Z_i$ is an equivalent form of the two-sample $t$-statistic $t_i$ in (1.1).

Once again, however, there are difficulties with this as a test for column-wise independence. There is no question that the $Z_i$'s are overdispersed compared to the theoretical value $\tau = 1$. But problems other than column dependence can cause overdispersion, in particular unobserved covariate differences between subjects in the two samples [Efron (2004, 2008)].

The statistic $S = w'\hat{\Delta}w$ in (5.15) does not depend upon the order of the columns of $X$ within each of the two samples, answering objection (2) against permutation tests, but it is the only such choice for a two-sample situation. Other $w$'s might yield interesting results. The version of (5.15) comparing the first 22 healthy Cardio subjects with the second 22 provided the spectacular value $\hat{\tau} = 1.87$, and here the "unobserved covariate" objection has less force.

Now, however, the test statistic depends on the order of the columns within the healthy subjects' matrix, reviving objection (2). Again we might want to check a catalog of possible $w$ vectors $w_1, w_2, \ldots, w_H$, leading back to test statistic

$$(5.17) \qquad S_B = \sum_h w'_h \hat{\Delta} w_h = \mathrm{tr}(\hat{\Delta}B) \qquad \left(B = \sum_h w_h w'_h\right),$$

as in (2.10), the only difference being that the null distribution of $\hat{\Delta}$ now involves normal theory rather than permutations. Remark 6.9 shows that the null first and second moments of $S_B$ are similar to (5.12),

$$(5.18) \qquad S_B \underset{H_0}{\sim} \left(\mathrm{tr}(B), \frac{2}{\tilde{m}} \mathrm{tr}(B^2)\right).$$

In summary, normal-theory methods are interesting and promising, but are not yet proven competitors for the permutation tests of Section 2.

**6. Remarks.** This section presents some brief remarks and details supplementing the previous material.

REMARK 6.1 (The constant $c_2$). The variance constant $c_2$ in Theorem 1 (3.3) can be expressed as

$$(6.1) \qquad c_2 = \frac{K}{(mn)^2}\left[\bar{e}^2 + \sum_{k=1}^{K}(e_k - \bar{e})^2\right] \qquad \left(\bar{e} \equiv \sum_{1}^{K} e_k/K\right),$$

so that $c_2 \geq K(\bar{e}/mn)^2$, with equality only if the eigenvalues $e_k$ are equal. In the doubly standardized case $\bar{e} = mn/K$, giving

$$(6.2) \qquad\qquad\qquad\qquad c_2 \geq 1/K,$$

where $K$ is the rank of $X$.

REMARK 6.2 (Permutation invariance). If the columns of $X$ are i.i.d. observations from a distribution on $\mathbb{R}^m$, then the distribution of $X$ is invariant under permutations: $X\boldsymbol{\pi} \sim X$ for any $n \times n$ permutation matrix $\boldsymbol{\pi}$. Now suppose $\widetilde{X} = L(X)$, where $L$ performs the same operation on each column of $X$, for example, replacing each column by its normal scores vector. Then

$$(6.3) \qquad \widetilde{X}\boldsymbol{\pi} = L(X)\boldsymbol{\pi} = L(X\boldsymbol{\pi}) \sim L(X) = \widetilde{X},$$

showing that $\widetilde{X}$ is permutation invariant.

Similarly, suppose $\widetilde{X} = R(X)$, $R$ performing the same operation $\widetilde{X}_i = r(X_i)$ on each row of $X$, where now we require $r(x)\boldsymbol{\pi} = r(x\boldsymbol{\pi})$ for all $n$-vectors $x$. The same argument as (6.3) demonstrates that $\widetilde{X}$ is still permutation invariant. Iterating row and column standardizations as in Table 1 then shows that if the original data matrix $X$ is permutation invariant, so is its doubly standardized version.

REMARK 6.3 (Covariances after demeaning). Suppose that $X$ is normally distributed, with covariances $\Sigma \otimes \Delta$ (4.2), all columns having the same expectation vector $\mu$. Let $\widetilde{X}$ be the demeaned matrix obtained by subtracting all the row and column means of $X$. Then

$$(6.4) \qquad\qquad \widetilde{X} \sim \mathcal{N}_{m,n}(0, \widetilde{\Sigma} \otimes \widetilde{\Delta}),$$

where

$$(6.5) \qquad \widetilde{\Delta}_{jj'} = \Delta_{jj'} - \Delta_{\cdot j'} - \Delta_{j\cdot} + \Delta_{\cdot\cdot},$$

dots indicating averaging over the missing subscripts, and similarly for $\widetilde{\Sigma}$. This shows that demeaning tends to reduce covariances by recentering them around 0.

Table 1

*Successive row and column standardizations of the $20{,}426 \times 44$ matrix of healthy Cardio subjects. "Col" empirical standard deviation of $\widehat{Cor}_{jj'}$, $j < j'$; "Eig" $\hat{\alpha}$ from (3.6); "Row" from 1% sample of $\widehat{cor}_{ii'}$ values, adjusted for overdispersion (6.6), sampling standard error 0.0034*

| | Col | Row | Eig | | Col | Row | Eig |
|---|---|---|---|---|---|---|---|
| Demeaned | 0.252 | 0.286 | 0.000 | Demeaned | 0.252 | 0.286 | 0.000 |
| Col | 0.252 | 0.249 | 0.251 | Row | 0.241 | 0.283 | 0.279 |
| Row | 0.242 | 0.255 | 0.246 | Col | 0.241 | 0.251 | 0.240 |
| Col | 0.242 | 0.241 | 0.242 | Row | 0.240 | 0.247 | 0.241 |
| Row | 0.241 | 0.246 | 0.235 | Col | 0.240 | 0.247 | 0.240 |
| Col | 0.241 | 0.244 | 0.241 | Row | 0.241 | 0.240 | 0.235 |
| Row | 0.241 | 0.245 | 0.234 | Col | 0.241 | 0.237 | 0.240 |
| Col | 0.241 | 0.238 | 0.241 | Row | 0.241 | 0.233 | 0.233 |

REMARK 6.4 (Standardization). A matrix $X$ is "column standardized" by individually subtracting the mean and dividing by the standard deviation of each column, and similarly for row standardization. Table 1 shows the effect of successive row and column standardizations on the $20{,}426 \times 44$ demeaned matrix of healthy Cardio subjects. Here "Col" is the empirical standard deviation of the 946 column-wise correlations $\widehat{Cor}_{jj'}$, $j < j'$; "Eig" is $\hat{\alpha}$ in (3.6); and "Row" is the empirical standard deviation "$\hat{\beta}$" of a 1% sample of the row correlations $\widehat{cor}_{ii'}$, but adjusted for overdispersion,

$$(6.6) \qquad \text{Row}^2 = \frac{n}{n-1}\left(\hat{\beta}^2 - \frac{1}{n-1}\right).$$

Sampling error of the Row entries is about $\pm 0.0034$.

The doubly standardized matrix $X$ used for Figure 3 was obtained after five successive column-row standardizations. This was excessive; the figure looked almost the same after two iterations. Other microarray examples converged equally rapidly, though small counterexamples can be constructed where double standardization is not possible.

Microarray analyzes usually begin with some form of column-wise standardization [Bolstad et al. (2003), Qiu, Klebanov and Yakovlev (2005b)], designed to negate "brightness" differences between the $n$ microarrays. In the same spirit, row standardization helps prevent incidental gene differences (e.g., very great or very small expression level variabilities) from obscuring the actual effects of interest. Standardization tends to reduce the apparent correlations as in Remark 6.3. Without standardization, the scatterplot in Figure 1 stretches out along the main diagonal, correlation 0.917, driven by genes with unusually large or small inherent expression levels.

REMARK 6.5 (Corrected estimates of the total correlation). Suppose that the true row correlations $cor_{ii'}$ have mean 0 and variance $\alpha^2$, as in (3.8) with $\overline{cor} = 0$, and that given $cor_{ii'}$, the usual estimate $\widehat{cor}_{ii'}$ has mean and variance

$$(6.7) \qquad \widehat{cor}_{i,i'} \doteq [cor_{ii'}, (1 - cor_{ii'}^2)^2/(n-3)],$$

(6.7) being a good normal-theory approximation [Johnson and Kotz 1970, Chapter 32]. Letting $\bar{\alpha}^2$ be the empirical variance of the $\widehat{cor}_{ii'}$ values, a standard empirical Bayes derivation yields

$$(6.8) \qquad \hat{\alpha}^2 = A^2 - \frac{3}{n-5}A^4 \qquad \left[ A^2 = \frac{(n-3)\bar{\alpha}^2 - 1}{n-5} \right]$$

as an approximately unbiased estimate of $\alpha^2$. (If $\overline{cor}$ is not assumed to equal 0, a slightly more complicated formula applies.) Of course, $\hat{\alpha}^2 = 0$ if the right-hand side of (6.8) is negative.

Theorem 1 implies that $\bar{\alpha}^2$ nearly equals $c_2$, (3.3), in the doubly standardized situation. Formula (3.6), with, say,

$$(6.9) \qquad \tilde{\alpha}^2 = \frac{n}{n-1}\left( \bar{\alpha}^2 - \frac{1}{n-1} \right),$$

is not identical to (6.8), but provides an excellent approximation for values of $\bar{\alpha} \leq 0.5$: with $n = 44$ and $\bar{\alpha} = 0.283$ as in (3.6), $\hat{\alpha} = 0.2415$ while $\tilde{\alpha} = 0.2412$.

REMARK 6.6 (Column and row centerings). The column correlation mean $\hat{\mu} = -1/(n-1)$ in (3.6) is forced by the row-wise demeaning $\sum_j x_{ij} = 0$, (2.1), centering the solid histogram in the right panel of Figure 3 at $-0.023$. With $m = 20{,}426$, the corresponding center for the line histogram is nearly 0, and the difference in the two centerings is noticeable. The dashed density curve in Figure 4, and the corresponding $p$-values for the FDR analysis, were shifted 0.023 units leftward.

REMARK 6.7 (The total correlation $\alpha$). The total correlation $\alpha$, which plays a key role in Theorem 2, (4.9), also is the central parameter of the theory developed in Efron (2007a). Equations (3.15)–(3.16) there are equivalent to (5.12) here. In both papers, $\alpha$ has the very convenient feature of summarizing the effects of an enormous $m \times m$ correlation matrix $\Sigma$ in a single number.

REMARK 6.8 [$\Sigma$ for simulation (5.11)]. The $X^*$ simulation used in Figure 5 began with $m \times n$ matrix $Y = (y_{ij})$,

$$(6.10) \quad y_{ij} = c_{Ij} + e_{ij}, \qquad \begin{cases} e_{ij} \sim \mathcal{N}(0, 1), \\ c_{Ij} \sim \mathcal{N}(0, \gamma^2) \end{cases} \qquad \text{(all independent)},$$

where $I = 1, 2, 3, 4, 5$ as $i$ is in the first, second, ..., last fifth of 1 through $m$; $Y$ was then column standardized to give $X^*$, so that $\hat\Sigma$ had a block form, with large positive correlations (about 0.61) in the $(m/5) \times (m/5)$ diagonal blocks. The choice $\gamma = 1.23$ was required to yield $\alpha = 0.241$.

REMARK 6.9 (Bilinear statistics). Since $\widetilde\Delta \sim (\Delta, \Delta^{(2)}/\tilde m)$ (4.8), it is clear that $E\{\tilde\tau^2\} = \tau^2$ in Corollary (4.20). The variance calculation proceeds as in Theorem 2:

$$
\begin{aligned}
\mathrm{var}\{\tilde\tau^2\} &= \sum_{jk}\sum_{lh} \Delta^{(2)}_{jk,lh} w_j w_k w_l w_h / \tilde m \\[2mm]
&= \sum_{jk}\sum_{lh} [\Delta_{jl}\Delta_{kh} + \Delta_{jh}\Delta_{kl}] w_j w_k w_l w_h / \tilde m \\[2mm]
(6.11)\qquad &= \left[\sum_{jl}\sum_{kh}(\Delta_{jl}w_j w_l)(\Delta_{kh}w_k w_h) + \sum_{jh}\sum_{kl}(\Delta_{jh}w_j w_h)(\Delta_{kl}w_k w_l)\right]\Big/ \tilde m \\[2mm]
&= 2\left(\sum_{jk}\Delta_{jk}w_j w_l\right)^2 \Big/ \tilde m = 2\tau^4/\tilde m.
\end{aligned}
$$

The verification of (5.18) is the same, except with element $b_{jk}$ of $B$ replacing $w_j w_k$ above, $b_{lh}$ replacing $w_l w_h$, etc.

# REFERENCES

ANDERSON, T. W. (2003). *An Introduction to Multivariate Statistical Analysis*, 3rd ed. Wiley, New York. MR1990662

BENJAMINI, Y. and HOCHBERG, Y. (1995). Controlling the false discovery rate: A practical and powerful approach to multiple testing. *J. Roy. Statist. Soc. Ser. B* **57** 289–300. MR1325392

BOLSTAD, B. M., IRIZARRY, R. A., ÅSTRAND, M. and SPEED, T. P. (2003). Comparison of normalization methods for high density oligonucleotide array data based on variance and bias. *Bioinformatics* **19** 185–193. Available at http://web.mit.edu/biomicro/education/RMA.pdf.

CALLOW, M., DUDOIT, S., GONG, E., SPEED, T. and RUBIN, E. (2000). Microarray expression profiling identifies genes with altered expression in HDL-deficient mice. *Genome Research* **10** 2022–2029.

EFRON, B. (2004). Large-scale simultaneous hypothesis testing: The choice of a null hypothesis. *J. Amer. Statist. Assoc.* **99** 96–104. MR2054289

EFRON, B. (2007a). Correlation and large-scale simultaneous significance testing. *J. Amer. Statist. Assoc.* **102** 93–103. MR2293302

EFRON, B. (2007b). Size, power, and false discovery rates. *Ann. Statist.* **35** 1351–1377. MR2351089

EFRON, B. (2008). Microarrays, empirical Bayes, and the two-groups model (with discussion and rejoinder). *Statist. Sci.* **23** 1–47. MR2431866

JOHNSON, D. E. and GRAYBILL, F. A. (1972). An analysis of a two-way model with interaction and no replication. *J. Amer. Statist. Assoc.* **67** 862–868. MR0400566

Johnson, N. L. and Kotz, S. (1970). *Continuous Univariate Distributions* **1**. Houghton Mifflin Company, Boston.

Mardia, K., Kent, J. and Bibby, J. (1979). *Multivariate Analysis*. Academic Press, London/San Diego.

Owen, A. B. (2005). Variance of the number of false discoveries. *J. Roy. Statist. Soc. Ser. B* **67** 411–426. MR2155346

Qiu, X., Brooks, A. I., Klebanov, L. and Yakovlev, A. (2005). The effects of normalization on the correlation structure of microarray data. *BMC Bioinformatics* **6** 120. Available at http://www.biomedcentral.com/1471-2105/6/120.

Qiu, X., Klebanov, L. and Yakovlev, A. (2005). Correlation between gene expression levels and limitations of the empirical Bayes methodology for finding differentially expressed genes. *Statist. Appl. Genet. Mol. Bio.* **4**, article 34. Available at http://www.bepress.com/sagmb/vol4/iss1/art34. MR2183944

Singh, D., Febbo, P. G., Ross, K., Jackson, D. G., Manola, J., Ladd, C., Tamayo, P., Renshaw, A. A., D'Amico, A. V., Richie, J. P., Lander, E. S., Loda, M., Kantoff, P. W., Golub, T. R. and Sellers, W. R. (2002). Gene expression correlates of clinical prostate cancer behavior. *Cancer Cell* **1** 203–209.

Tusher, V. G., Tibshirani, R. and Chu, G. (2001). Significance analysis of microarrays applied to the ionizing radiation response. *Proc. Nat. Acad. Sci. USA* **98** 5116–5121. Available at http://www.pnas.org/cgi/content/full/98/9/5116.

Department of Statistics
Sequoia Hall
390 Serra Mall
Stanford, California 94305-4065
USA
E-mail: brad@stat.stanford.edu